\documentclass[twocolumn,pre,showpacs]{revtex4-1}
\usepackage{epsfig}
\usepackage{amsmath}
\usepackage{amsfonts}

\begin{document}

\title{Quantum phase transitions in an interacting atom-molecule boson model}

\author{G. Santos$^1$, A. Foerster$^2$, J. Links$^3$, E. Mattei$^2$, S.R. Dahmen$^2$}
\address{$^1$ Departamento de F\'{\i}sica,
Universidade Federal de Sergipe, S\~ao Crist\'ov\~ao, SE, Brazil}
\address{$^2$ Instituto de F\'{\i}sica, Universidade Federal do Rio Grande do Sul,
Porto Alegre, RS, Brazil}
\address{$^3$ Department of Mathematics,
University of Queensland, Brisbane, QLD, Australia}

\begin{abstract}
We study the quantum phase transitions of a model that describes the interconversion of interacting bosonic atoms and 
 molecules. Using a classical analysis, we identify a threshold coupling line separating a molecular  phase and a mixed phase. Through studies of the energy gap, von Neumann entanglement entropy, and fidelity,  
we give evidence that this line is associated to a boundary line in the ground-state phase diagram of the quantum system. 

\end{abstract}



\maketitle
\parskip 2mm

\section{Introduction}

A Quantum Phase Transition (QPT) is characterized by a change 
in the properties of the ground state of a system (i.e. at zero temperature)
as some parameter is varied across a critical point
 \cite{Sachdev,Sondhi}. This parameter can be an external magnetic field as in the quantum Hall effect or
superconducting materials or, as in a Bose-Einstein condensate, a change in the
the $s$-wave scattering amplitude or external fields to produce an atom-molecule condensate \cite{Zoller}.  
The change in the nature of the ground state is typically identified by non-analyticity of some quantity, such as the ground-state energy or a correlation function. A common approach for studying QPTs is to use concepts such as order parameters and symmetry breaking {\it a la} Landau-Ginzburg, in analogy with thermal phase transitions. 
However in many cases of interest order
parameters are difficult to identify. More recently, in the wake of quantum information theory, 
it has been realized that one can employ alternative
concepts such as entanglement \cite{Review} and fidelity \cite{Zanardi,huan} as means of identifying different phases.

Bosonic models are useful testbeds for investigating QPTs in that they accommodate large particle numbers with few degrees of freedom, in contrast to fermionic systems, which simplifies the analysis.  
In this paper we study QPTs for an interacting atom-molecule boson model.
The model does not possess any symmetries which  might give rise to symmetry-breaking order. Using other methods, we will establish a line of QPTs in the ground-state  phase diagram.  
This line is first identified via phase space bifurcations in a classical analysis of the system, adapting a correspondence that has been put forth in \cite{Milburn2} relating bifurcations in classical systems with entanglement entropy of quantum systems. We then confirm that this line 
is associated with quantum phase transitions of the quantum system through studies of the energy gap, entanglement entropy, and fidelity.   
Even though a QPT is 
only rigorously defined in the thermodynamic limit $ N \rightarrow \infty $, where $N$ is the number or particles 
in the system, we will show that the aforementioned concepts do respond
to changes in the ground state of the system for finite $N$ and strongly indicate the presence
of QPTs (cf. \cite{iachello}).  

We consider the model for interacting 
 atomic and molecular bosons  described by the following Hamiltonian \cite{jonreview}
\begin{eqnarray}
H&=&U_aN_a^2 + U_bN_b^2 +U_{ab}N_aN_b + \mu_aN_a +\mu_bN_b \nonumber \\ 
&&+\Omega(a^{\dag}a^{\dag}b +b^{\dag}aa).
\label{hamq1}
\end{eqnarray}

\noindent where $a^{\dagger}$ is the creation operator for an atomic mode and 
$b^{\dagger}$ is the creation operator for a molecular mode. The parameters $\mu_i$ are chemical 
 potentials for species $i$ and $\Omega$ is the amplitude for the interconversion of atoms and molecules. 
The Hamiltonian commutes
with the total atom number $N=N_a+2N_b$,
where $N_a=a^{\dagger}a$ and $N_b=b^{\dagger}b$. Thus the Hamiltonian can be block diagonalized into sectors on which $N$ takes a constant value. Hereafter we consider the action of the Hamiltonian to be restricted to such a sector and treat $N$ to be a scalar variable. 
Notice that the change of variable $\Omega\rightarrow -\Omega$ is equivalent to the unitary transformation
\begin{equation}
 b\rightarrow - b. 
\label{trans}
\end{equation} 
\noindent 
The parameters $U_{j}$ describe $s$-wave scattering, taking into 
account the atom-atom ($U_{a}$), atom-molecule ($U_{ab}$) and molecule-molecule ($U_{b}$) interactions. 

In the no-scattering limit $U_{a}=U_{ab}=U_{b}=0$, the model (\ref{hamq1}) has been studied using a
variety of methods \cite{Vardi1,Milburn1,Jon1,ivkw}, and it was found to
undergo a quantum phase transition when  $\mu_a/\Omega=\sqrt{2N}$.
This was confirmed analytically through a Bethe Ansatz solution~\cite{claire}. 
However in the experimental context the $s$-wave 
scattering interactions play a significant role \cite{Inouye,Heinzen,cusak,Thierry,xll}. 
It was shown in \cite{GSantos1,iw,chin} that for the  
general model (\ref{hamq1}) the inclusion of these scattering terms has non-trivial consequences for the physical behaviour.
In this paper we re-examine the QPT  found in the no-scattering limit, 
showing how the concepts 
of quantum information, entanglement and fidelity are related to it. 
Moreover, we also investigate the effect of the $s$-wave scattering 
parameters in the QPTs, establishing a connection 
between the QPTs and the bifurcation line
associated to the global minimum of the Hamiltonian (\ref{hamq1})
in the parameter space of the classical system.

The paper is organized as follows: in section 2 we give an outline of the classical analysis of 
the model (\ref{hamq1}) including the fixed point bifurcations that occur in the classical phase space. In section 3 we investigate the behaviour of the energy gap, entanglement, and fidelity 
to identify QPTs. We also establish a connection between these QPTs and a 
bifurcation line of the global minimum in the parameter space of the classical system.
Our conclusions are stated in section 4. 

\section{Classical analysis}

We first motivate undertaking a classical analysis to gain insights into the existence of QPTs in quantum systems. Recall the proposition in \cite{Milburn2} that, if for a bipartite quantum system there exists a supercritical pitchfork bifurcation of the global minimum at some coupling parameter in the phase space of the analogous classical system, then the quantum system will take a maximum value of the ground-state entanglement entropy at that coupling. This proposition holds true for attractive bosons in a double-well potential, within a two-mode approximation. The double-well model admits a QPT associated with symmetry-breaking as shown in \cite{clmz98}. The critical coupling coincides with the supercritical bifurcation of the global minimum in phase space \cite{ks02}, and also coincides with the point where the entanglement entropy is maximal \cite{pan}.

To investigate the extent to which these ideas extend to the present model we briefly recall the classical analysis of the Hamiltonian (\ref{hamq1}) as given in \cite{GSantos1}.
Let $N_j,\,\theta_j,\,j=a,\,b$ be 
quantum variables satisfying canonical commutation relations. We make a change of variables from the operators $j,\,j^\dagger,\,j=a,\,b$ via
$$j=\exp(i\theta_j)\sqrt{N_j},
~~~j^\dagger=\sqrt{N_j}\exp(-i\theta_j) $$ 
such that the canonical commutation relations are preserved. Now define the variables
$$ z=\frac{1}{N}(N_a-2N_b),~~~\theta=\frac{N}{4}(2\theta_a-\theta_b),$$ 
such that $(z,\theta)$ are canonically conjugate variables.
In the classical limit where $N$ is large, but still finite, we may equivalently consider the transformed Hamiltonian \cite{GSantos1} 
\begin{equation}
H=\lambda z^2 +2 \alpha z +\beta   
+\sqrt{2(1-z)}(1+z) \cos\left(\frac{4\theta}{N}\right)
\label{ham2}
\end{equation}
where
\begin{eqnarray}
 \lambda &=& \frac{\sqrt{2N}}{\Omega}\left(\frac{U_{a}}{2}
-\frac{U_{ab}}{4}+\frac{U_{b}}{8}
\right)  \\
\alpha &=&\frac{\sqrt{2N}}{\Omega}\left(\frac{U_{a}}{2}
-\frac{U_{b}}{8} + \frac{\mu_a}{2N}-\frac{\mu_b}{4N}\right)   \\
\beta &=& \frac{\sqrt{2N}}{\Omega}\left(\frac{U_{a}}{2}
+\frac{U_{ab}}{4}+\frac{U_{b}}{8}+\frac{\mu_a}{N}+\frac{\mu_b}{2N}
\right).
\label{param}   
\end{eqnarray} 

We note that the unitary transformation (\ref{trans})
is equivalent to $\theta \rightarrow \theta +{N\pi}/{4}$. 
Hereafter we restrict our attentions to the ``repulsive'' case $\lambda \geq 0$. 

\begin{figure}
\centerline{\epsfig{figure=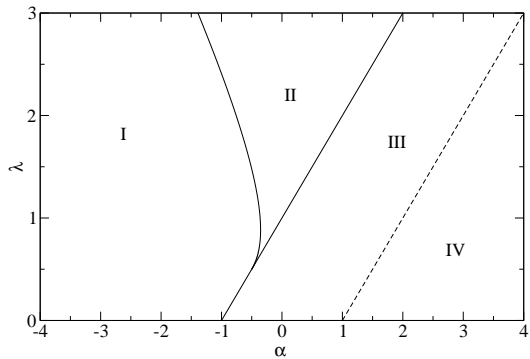,width=0.80\linewidth}}
\caption{Parameter space diagram identifying the different types of solution for  
equation (\ref{fixed}). In region I there are no solutions for $z$ when $\theta = 0$, and one solution for 
$z$ when $\theta = {N\pi}/{4}$. In region II there are two solutions for $z$ when $\theta = 0$, and one solution for 
$z$ when $\theta = {N\pi}/{4}$. In region III there exists one solution for $z$ when $\theta = 0$,  one solution for 
$z$ when $\theta = {N\pi}/{4}$, and two solutions for $\theta$ when $z=-1$. In region IV there is one solution for $z$ 
when $\theta = 0$,  
and no solution for $z$ when  
$\theta = {N\pi}/{4}$. In this region the global minima occur at $z=-1$ and all values of $\theta$. The boundary separating regions I and II of the region III is given by $\lambda=\alpha+1$, while 
the equation $\lambda =\alpha-1$ separates the region III of the region IV.
The boundary between regions I and II has been obtained numerically.}
\label{fig3}
\end{figure} 

We now regard (\ref{ham2}) as a classical Hamiltonian and 
investigate the global minima of the system. The first step is to 
find Hamilton's equations of motion which yields
\begin{eqnarray} 
\frac{dz}{dt}&=&\frac{\partial H}{\partial \theta}=-\frac{4}{N}\sqrt{2(1-z)}  
(1+z)  \sin\left(\frac{4\theta}{N}\right),  \label{de1} \\
-\frac{d\theta}{dt}&=&\frac{\partial H}{\partial z} =2\lambda z +2\alpha 
+\frac{1-3z}{\sqrt{2(1-z)}} \cos\left(\frac{4\theta}{N}\right).
\label{de2}  
\end{eqnarray}
Within the interior of the compact phase space the fixed points of the system are determined by the condition
\begin{equation}
\frac{dz}{dt}=\frac{d\theta}{dt}=0. 
\label{fixed}
\end{equation}

Extremal points may also occur on the boundaries $z=1,\,z=-1$. The bifurcations of the fixed points
divide the coupling parameter space into different regions, as shown in Fig. \ref{fig3}. 
Of these bifurcations, only the boundary separating regions III and IV (the line $\lambda = \alpha-1$) is associated with 
a qualitative change of the global minimum of the Hamiltonian (\ref{ham2}). In the regions I, II, III qualitative changes are associated with either saddle points or 
maxima (see \cite{GSantos1} for details). For these regions the minimum of (\ref{ham2}) occurs at $\theta=N\pi/4$ with $z>-1$. In region IV there is a line of global minima for (\ref{ham2}) at the boundary $z=-1$ and for all values of $\theta$. These global minima do not satisfy (\ref{fixed}).

From the above we see that there is a bifurcation of the phase space minimum from region III, where there is a unique
minimum, to region IV, where there is a line of minima. The bifurcation line in the parameter space, which is given by
$\lambda=\alpha-1$, is not of a supercritical pitchfork type discussed in \cite{Milburn2}. Consequently, we should not
expect the ground-state entanglement entropy to be maximal on this bifurcation line. In fact it has been shown in
\cite{Milburn1} that for the point $(\lambda,\,\alpha)=(0,1)$ on this line the entanglement entropy is {\it not}
maximal. Nonetheless, we will establish below that the bifurcation line $\lambda=\alpha-1$
is still associated to a line of QPTs for the quantum Hamiltonian (\ref{hamq1}).

\section{Quantum phase transitions}

\subsection{Energy gap}
\begin{figure}
\centerline{\epsfig{figure=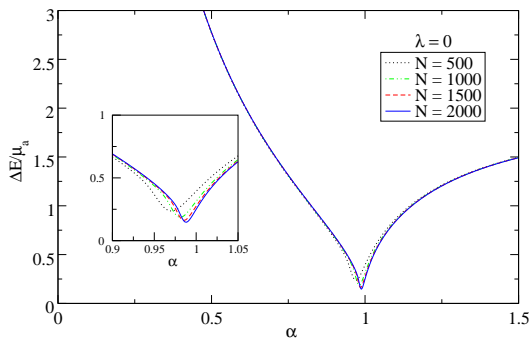,width=0.8\linewidth}}
\caption{(Color online) Dimensionless energy gap between the first excited state and the ground state as a function of 
$\alpha = \mu_a / \Omega\sqrt{2N}$ for different values of $N$ and $\lambda = 0$. Here $\Omega = 1$, $\mu_b = 0$ and $U_a =U_b/4= 0.25$.}
\label{gap1}
\end{figure}

\begin{figure}
\centerline{\epsfig{figure=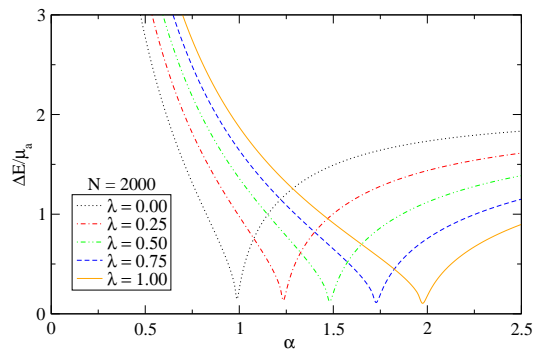,width=0.8\linewidth}}
\caption{(Color online) Dimensionless energy gap between the first excited state and the ground state as a function of 
$\alpha = \mu_a / \Omega\sqrt{2N}$ for different values of $\lambda$ and $N=2000$. 
Values of $\Omega$, $\mu_b$ and $U_a$, $U_b$ are the same as in the previous figure. These results indicate that the minimal values lie approximately on the line $\lambda=\alpha-1$.}
\label{gap2}
\end{figure}
\vspace{1.0cm}
We begin our analysis by considering the dimensionless energy gap between the first excited state and the ground state, $\Delta E/\mu_a$. 
Using numerical diagonalization of the Hamiltonian, in Fig. \ref{gap1} we plot the scaled gap $\Delta E/\mu_a$,  as a function of the coupling  
$\alpha$,  for $\lambda = 0$ and different values of $N$.
We observe that as $N$ increases the dimensionless gap decreases and the coupling approaches
the value $\alpha = 1$. 
Fig. \ref{gap2} shows similar results 
for fixed $N=2000$ and varying $\lambda$. We
observe that the occurrence of the minima of the gap, determining the QPT, fit well with the predicted 
boundary separating regions III and IV given by $\lambda =\alpha-1$. This is the first evidence suggesting that a line of QPTs occurs. 
Note that in both graphs the scaled gap goes to zero at a single point, rather than over an interval, of the coupling $\alpha$. This is indicative of the fact that there is no phase where the ground state is degenerate, which would be a requirement for the existence of a broken-symmetry phase. 

\begin{figure}
\centerline{\epsfig{figure=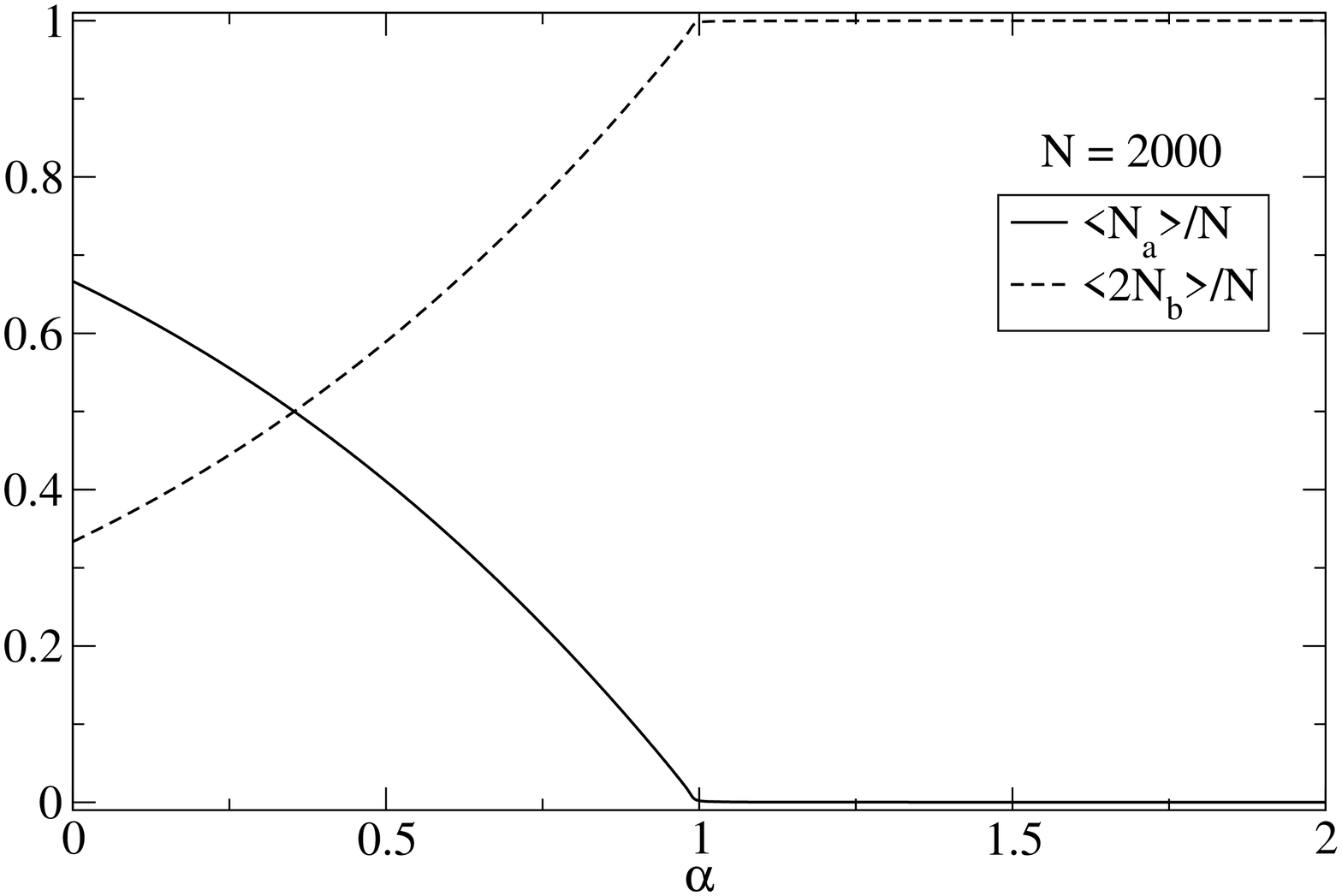,width=0.8\linewidth}}
\caption{Ground-state expectation values for the atomic number fraction $N_a/N$ and  the molecular number fraction $N_b/N$
as a function of 
$\alpha =\mu_a / \Omega\sqrt{2N} $ for $\lambda = 0$ and $N=2000$.
  We have set $\Omega = 1$, $\mu_b = 0$ and $U_a =U_b/4= 0.25$. There is a sharp transition at $\alpha=1$.}
\label{qpt_ab}
\end{figure}

To better understand the physical meaning of the QPTs, we depict in Fig. 
\ref{qpt_ab} the ground-state expectation value of the scaled atomic 
number operator (solid line) and the scaled molecular number operator (dashed line) as function of $\alpha $
for $N=2000$ and $\lambda =0$. The average value for the number of atoms 
decreases while the average number of molecules increases as $\alpha$ increases. 
For $\alpha >1$, the average number of molecules is maximal,
therefore we can interpret this point $\alpha =1$ as the threshold coupling for the 
formation of a predominantly molecular state. This result is consistent with the classical analysis, whereby in region IV the global minima have $z=-1$ which corresponds to a molecular phase.

\subsection{Entanglement}

One may consider the atom-molecule model~(\ref{hamq1}) as a bipartite system of
two modes, $A$ and $B$. 
In this case, the standard measure of entanglement is the von 
Neumann entropy of the reduced density operator of either of the modes~\cite{Milburn1}.
The state of each mode is characterized by its occupation number. Using the fact that 
the total number of atoms $N$ is constant, a general state of the system 
can be written for even $N$ in terms of the Fock states by
\begin{equation}
 |\Psi\rangle = \sum_{n = 0}^N d_n|2n\rangle|N - n\rangle
\label{aaestado}
\end{equation}
\noindent where ${d_n}$ are complex numbers.

\begin{figure}
\centerline{\epsfig{figure=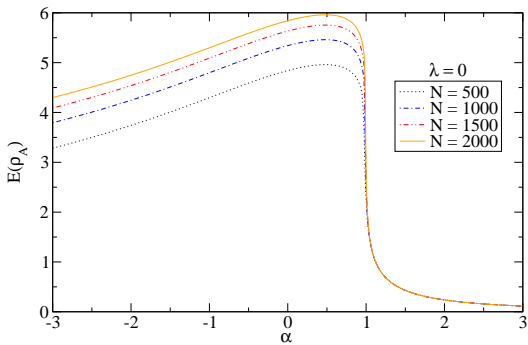,width=0.8\linewidth}}
\caption{(Color online) Entropy of entanglement of the ground state as a function of 
$\alpha = \mu_a / \Omega\sqrt{2N}$ for different values of $N$ and $\lambda = 0$. Here $\Omega = 1$, $\mu_b = 0$ and $U_a =U_b/4= 0.25$.}
\label{abentfig1}
\end{figure}

\begin{figure}
\centerline{\epsfig{figure=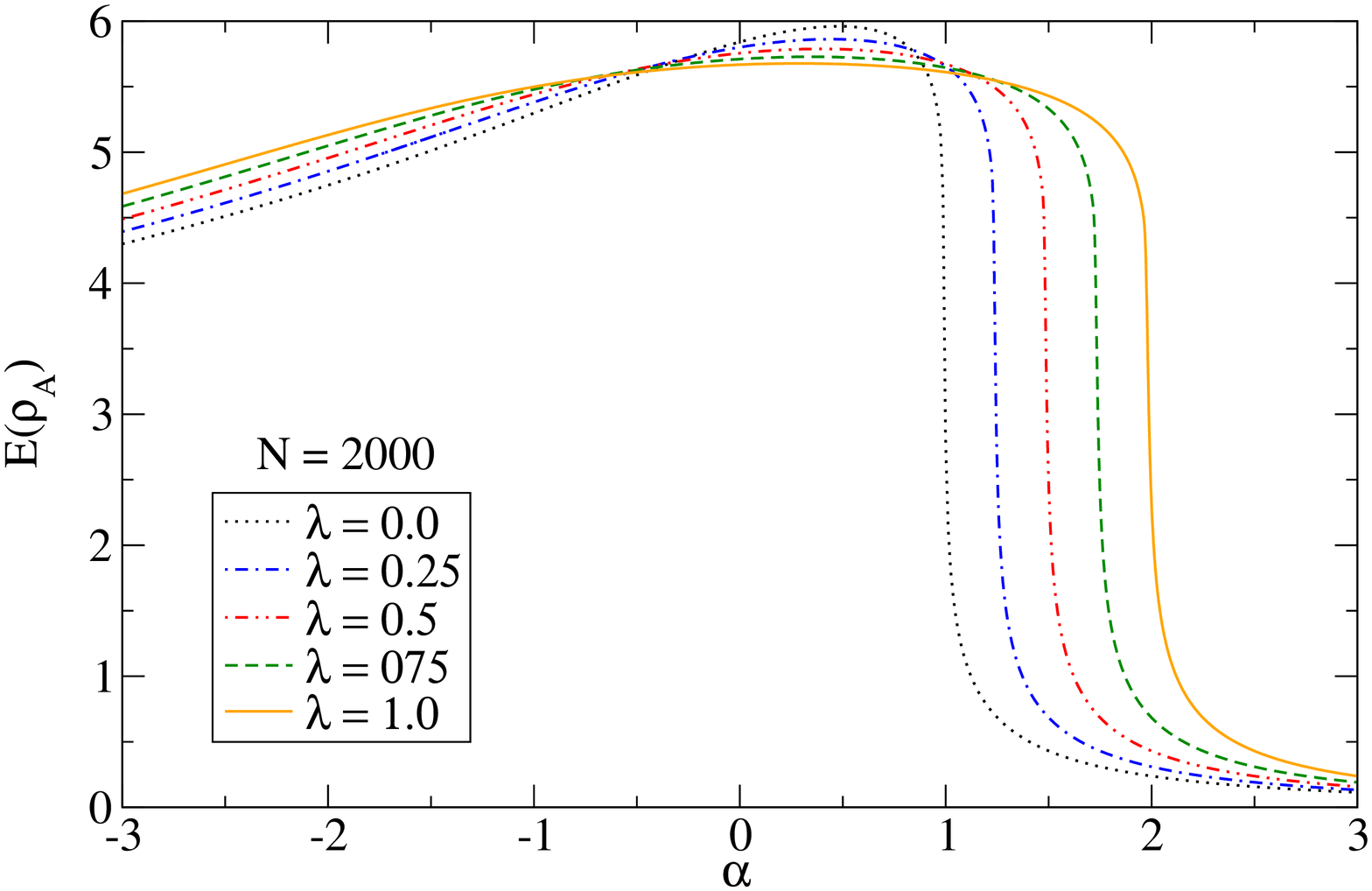,width=0.8\linewidth}}
\caption{(Color online) Entropy of entanglement of the ground state as a function of 
$\alpha = \mu_a / \Omega\sqrt{2N}$ for different values of $\lambda$ and $ N = 2000$. Here $\Omega = 1$, $\mu_b = 0$ and $U_a =U_b/4= 0.25$.}
\label{abentfig2}
\end{figure}
The density operator for state (\ref{aaestado}) is given by
 \begin{equation}
 \rho  =  |\Psi\rangle \langle\Psi| 
       =  \sum_{m,n = 0}^N d_m^*d_n|2m\rangle|N - m\rangle\langle 2n|\langle N - n|.
\label{aadensidade}
\end{equation}
\noindent Taking the partial trace with respect to the mode $B$ yields the reduced density
  operator for the mode $A$,
\begin{equation}
 \rho_A  =  \textrm{Tr}_B (\rho)
         =  \sum_{n = 0}^N |d_n|^2 |2n\rangle\langle 2n|.
\end{equation}
\noindent The entropy of entanglement of the ground-state of the system is given by
\begin{equation}
 E(\rho_A) = - \textrm{Tr}[\rho_A \textrm{log}_2
 (\rho_A)] = -\sum_{n = 0}^N |d_n|^2 \textrm{log}_2 (|d_n|^2).
\label{entrop1}
\end{equation}
Using the above expression (\ref{entrop1}) and the energy levels obtained through exact diagonalization of
the Hamiltonian (\ref{hamq1}) we plot in Figs. (\ref{abentfig1}) and (\ref{abentfig2}) the entropy of entanglement of the 
ground-state as a function of the coupling $\alpha$ for different values of $\alpha$ and $N$.

In Fig. (\ref{abentfig1}) we confirm that the entanglement entropy is not maximal at the threshold coupling $(\lambda, \alpha) = (0,1)$ determined from the classical analysis, in agreement with \cite{Milburn1}.
However we do observe that the entanglement entropy exhibits a sudden decrease close to $\alpha = 1$ that 
becomes more pronounced as the total number of
atoms $N$ increases.  
Fig. (\ref{abentfig2}) shows similar results for fixed $N=2000$ with varying $\lambda$. In this latter case we
see that the occurrence of the abrupt decay of the entropy fits with the predicted 
boundary separating regions III and IV given by $\lambda =\alpha-1$.

\begin{figure}
\centerline{\epsfig{figure=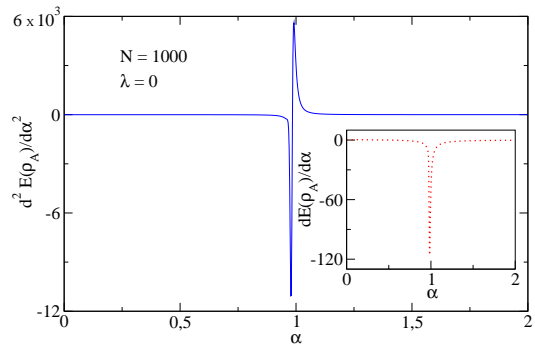,width=0.8\linewidth}}
\caption{(Color online) First (insert) and second  derivative  of the entropy of entanglement of the
 ground state as a function of $\alpha = \mu_a / \Omega\sqrt{2N}$ for $N=1000$ and $\lambda = 0$. Here $\Omega = 1$, $\mu_b = 0$ and $U_a =U_b/4= 0.25$.}
\label{entderiv2}
\end{figure}

We also depict in
Fig. (\ref{entderiv2}) the first
and second derivatives of the  ground state entanglement entropy as a function of $\alpha$.
Combined, Figs. (\ref{abentfig1}) and (\ref{entderiv2}) suggest a discontinuous behaviour in the limit $N\rightarrow\infty$, consistent with the existence of a QPT.  

\subsection{Fidelity}

\begin{figure}
\centerline{\epsfig{figure=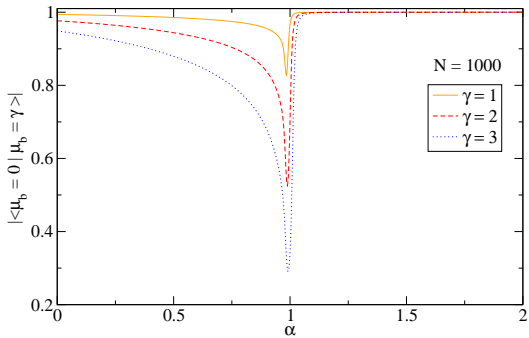,width=0.8\linewidth}}
\caption{(Color online) Fidelity of the ground state as a function of $\alpha = \mu_a / \Omega\sqrt{2N}$ for
  $\Omega=1$, $\lambda=0$, $N = 1000$ and $U_a =U_b/4= 0.25$. We keep $\mu_b=0$ for the 
 first state and vary $\mu_b=\gamma$ for the second state.}
\label{abfid1}
\end{figure}

Another possibility to investigate QPTs is through the behavior of the fidelity \cite{Zanardi,huan}. This concept is 
widely used in quantum information theory \cite{nielsen}.
The fidelity is defined as the modulus of the wavefunction overlap between two states
\begin{equation}
 \cal{F}(\psi,\phi)=|\langle \psi|\phi\rangle|. \nonumber
\end{equation}
In Fig. (\ref{abfid1}) we present the wavefunction overlap between two ground-states
corresponding  to the external parameter $\mu_b=0$ for one ground state 
and $\mu_b=\gamma$ for the other ground state. For $\gamma=0$ the states are
indistinguishable and there is no information about the QPT. When $\gamma$ increases 
the states become distinguishable and the fidelity has a minimum
at the same point where we find a QPT using the energy gap and entanglement entropy. 
For larger values of $\gamma$ the distinguishability increases but the minimum value remains in the
same position.

\begin{figure}
\centerline{\epsfig{figure=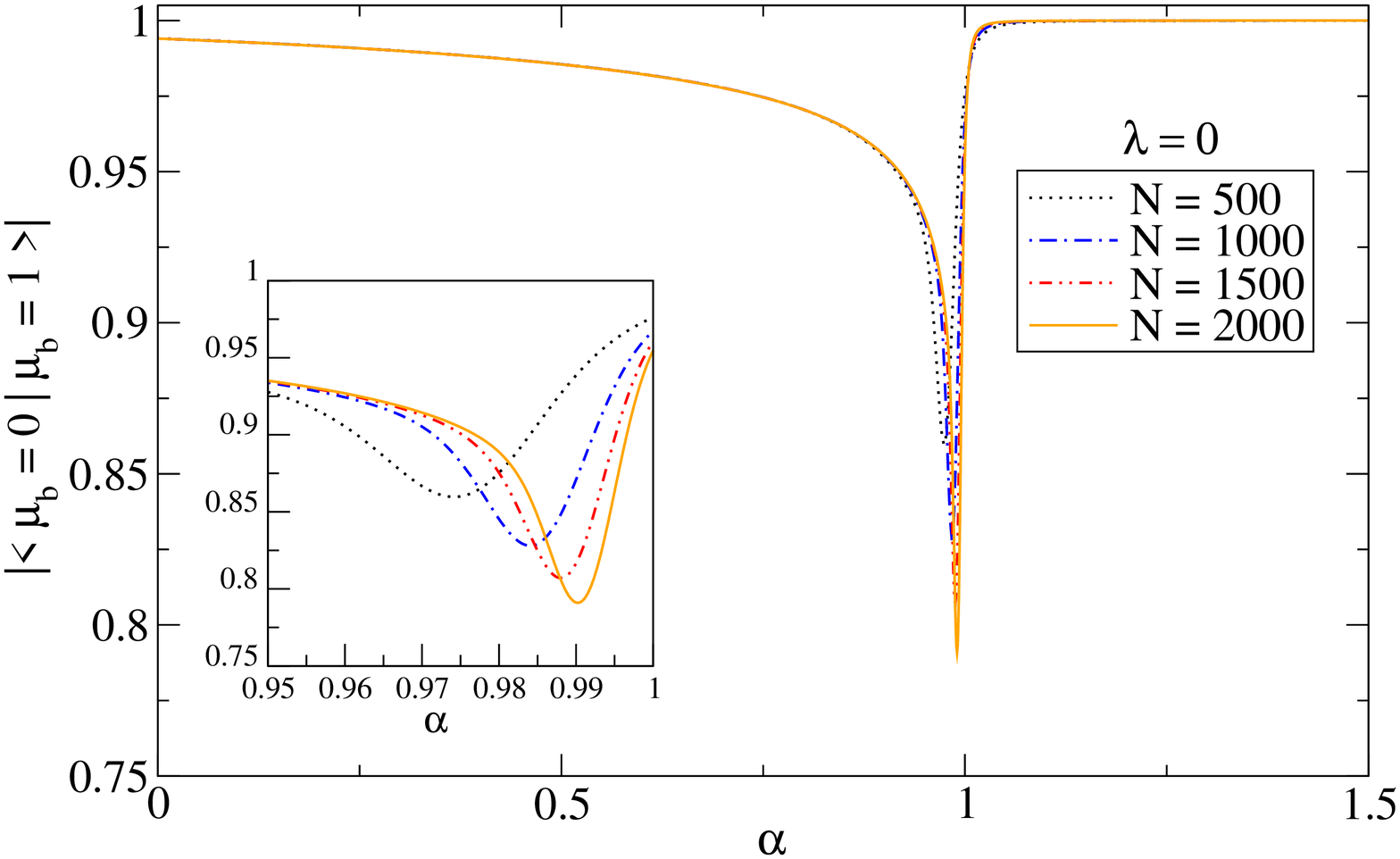,width=0.8\linewidth}}
\caption{(Color online) Fidelity of the ground state as a function of 
$\alpha = \mu_a / \Omega\sqrt{2N}$ for $\lambda=0$, $\mu_b=0$ (first 
 state) and  $\mu_b = 1$ (second state) with varying $N$.  Here $\Omega = 1$ and $U_a =U_b/4= 0.25$. The insert shows the minima close the critical point $\alpha = 1$.} 
\label{abfid2}
\end{figure}

\begin{figure}
\centerline{\epsfig{figure=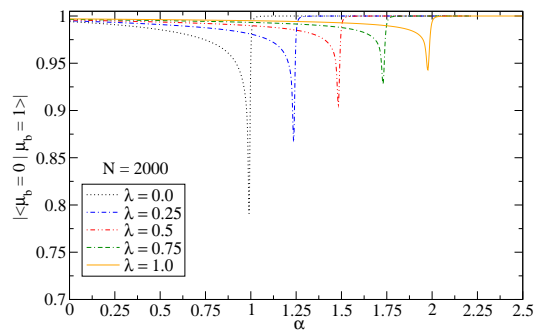,width=0.8\linewidth}}
\caption{(Color online) Fidelity of the ground state as a function of 
$\alpha = \mu_a / \Omega\sqrt{2N}$ for $N = 2000$, $\mu_b=0$ (first state) and  $\mu_b = 1$ (second state) and varying $\lambda$. Here $\Omega = 1$ and $U_a =U_b/4= 0.25$.}
\label{abfid3}
\end{figure}

Fig. (\ref{abfid2}) depicts the ground state fidelity  
  $|\langle \mu_b = 0|\mu_b=1 \rangle|$ for $\lambda=0$, $\gamma = 1$ and varying $N$. 
With increasing $N$
  the states become more distinguishable and the point where the minimum occurs moves towards $\alpha=1$. 
  Fig. (\ref{abfid3}) shows similar results for fixed $N=2000$ and varying $\lambda$. In this case we
also observe that the occurrence of the minima of the fidelity, determining the QPT, fit well with the predicted 
boundary line $\lambda =\alpha-1$.

\section{Summary}

We have used three different approaches to show that the interacting atom-molecule boson model 
(\ref{hamq1}) exhibits a line of QPTs that coincides with the boundary of the parameter space diagram 
determined by bifurcation line $\lambda = \alpha - 1$ of the global minimum of the Hamiltonian in the classical analysis.
First, the dimensionless energy gap is minimal on this line and indicates gapless excitations in the limit $N\rightarrow\infty$. Also, the derivatives of the entanglement entropy rapidly vary in the vicinity of this line. Finally, the fidelity approach shows that the states with $\mu_a=0$ and $\mu_a=\gamma$  
become distinguishable on this line,  where fidelity has a minimum. 
The parameter $\gamma$  changes the distinguishability of the states but does not change the value of the threshold coupling. 
All approaches indicate markedly different behaviours when the line $\lambda = \alpha - 1$ 
is crossed, giving a strong indicator towards the existence of a line of QPTs. These results show that the study of bifurcations in the phase space of classical systems  to identify quantum phases transitions may be applicable at a general level, independent of the nature of the bifurcations.

\subsection*{Acknowledgments}

A.F. would like to thank CNPq - Conselho Nacional de Desenvolvimento
Cient\'{\i}fico e Tecnol\'ogico for financial support. G. S. would like to thank Capes - Coordena\c{c}\~ao de Aperfei\c{c}oamento de Pessoal de N\'{\i}vel Superior for financial support. J.L. is supported by Australian Research Council Discovery Grant DP0663772.

\end{document}